\documentclass[a4paper,11pt]{article}
\usepackage{pos}

\title{Impact of charged and neutral Drell-Yan Asymmetries on precision measurements}

\author*[a]{Francesco Giuli}

\affiliation[a]{University of Rome ``Tor Vergata'' and INFN, Section of Rome 2,\\
  Via della Ricerca Scientifica 1, 00133, Rome, Italy}

\emailAdd{francesco.giuli@cern.ch}

\abstract{We study the impact of future measurements of lepton-charge and forward-backward asymmetries on PDF determination. These results have been obtained employing standard profiling procedures and the open-source platform \texttt{xFitter}. The potential of the combination of charged-current and neutral-current Drell-Yan asymmetries in regions of transverse and invariant masses near the $W/Z$ bosons peaks to improve the PDF uncertainties has also been explored.}

\FullConference{%
  *** The European Physical Society Conference on High Energy Physics (EPS-HEP2021), ***\\
  *** 26-30 July 2021 ***\\
  *** Online conference, jointly organized by Universität Hamburg and the research center DESY ***
}

\begin{document}
\maketitle

\begin{figure}[t!]
\begin{center}
\includegraphics[width=0.32\textwidth]{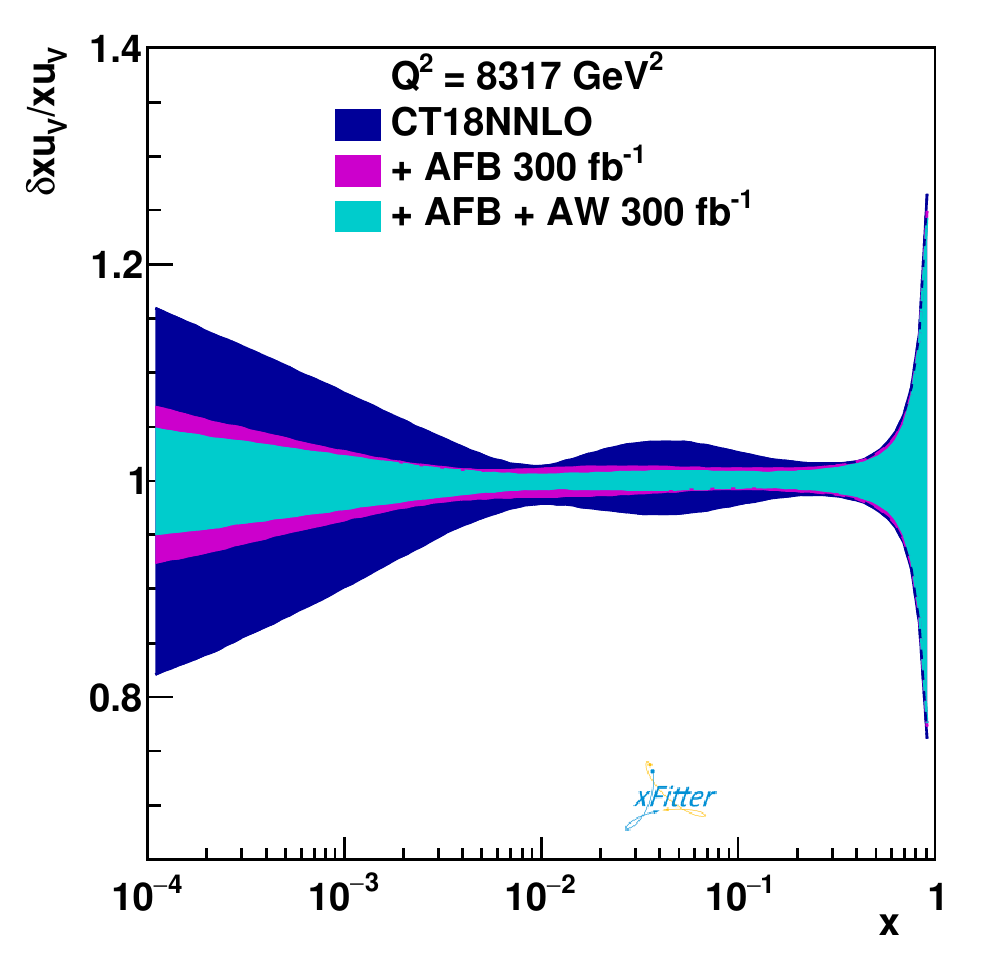}
\includegraphics[width=0.32\textwidth]{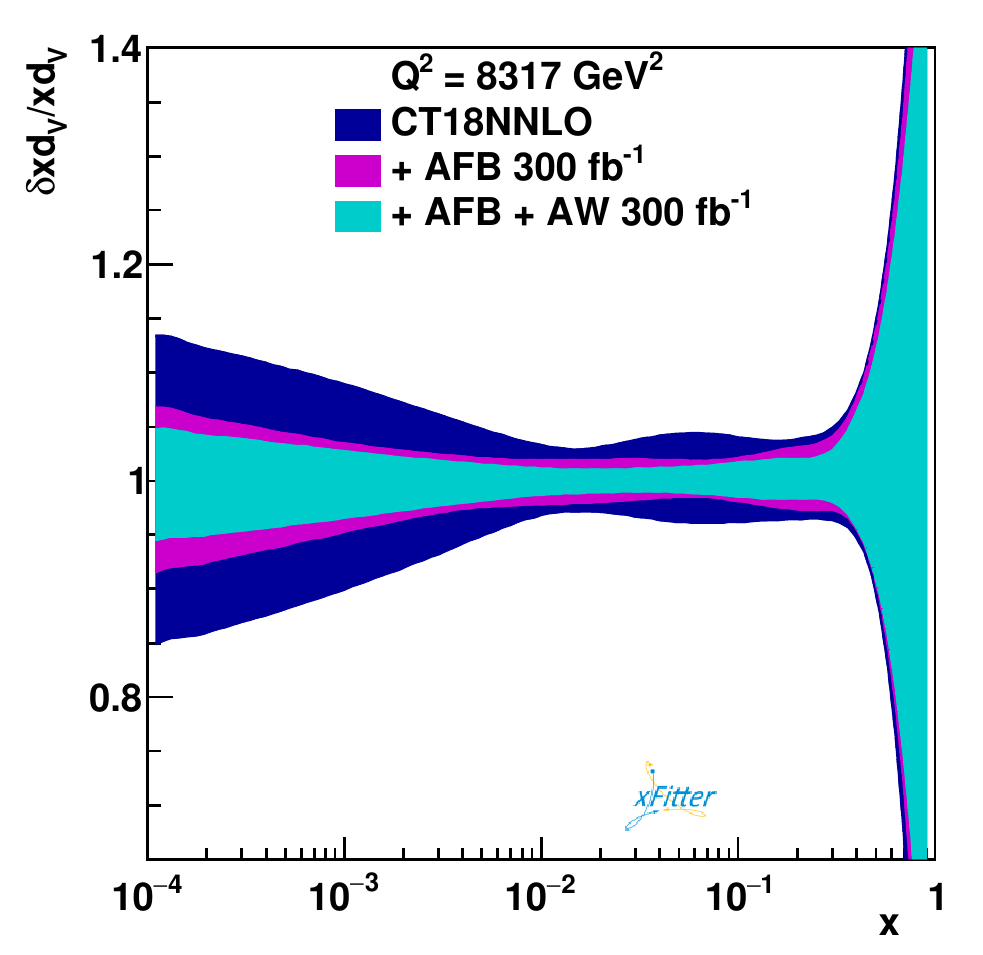}
\includegraphics[width=0.32\textwidth]{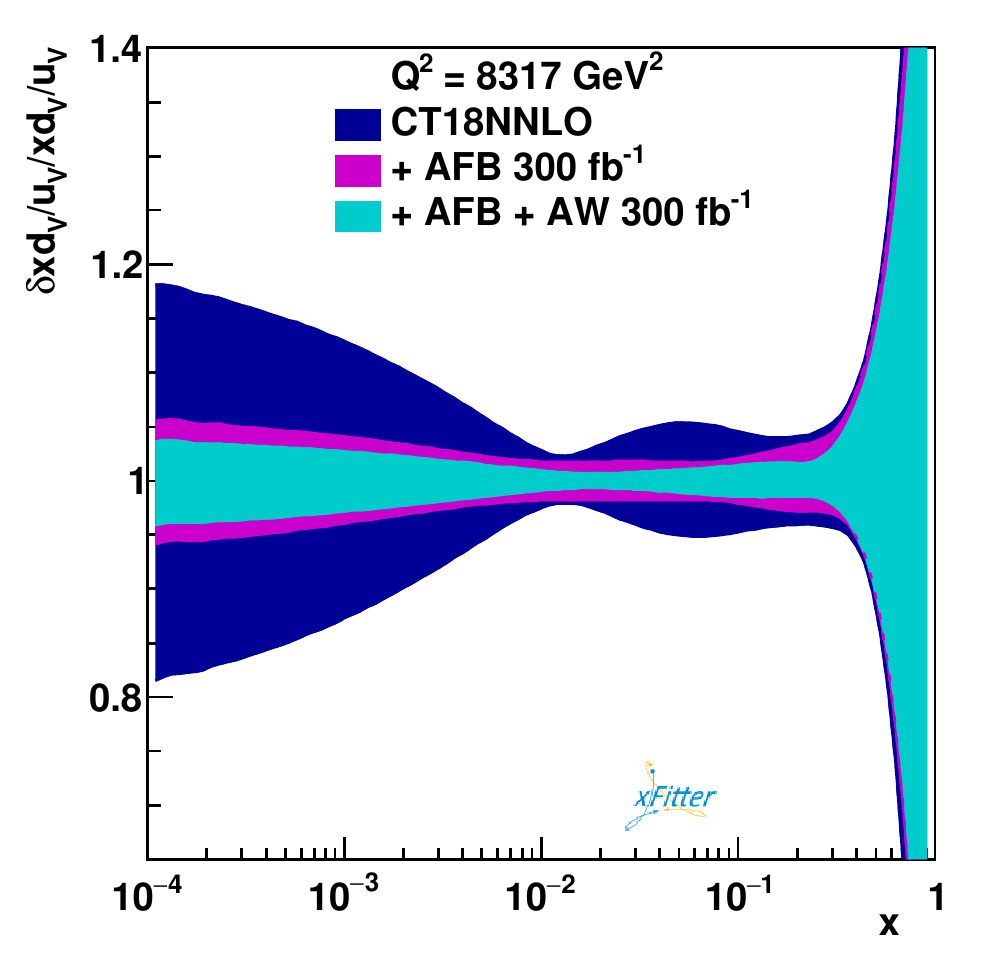}
\includegraphics[width=0.32\textwidth]{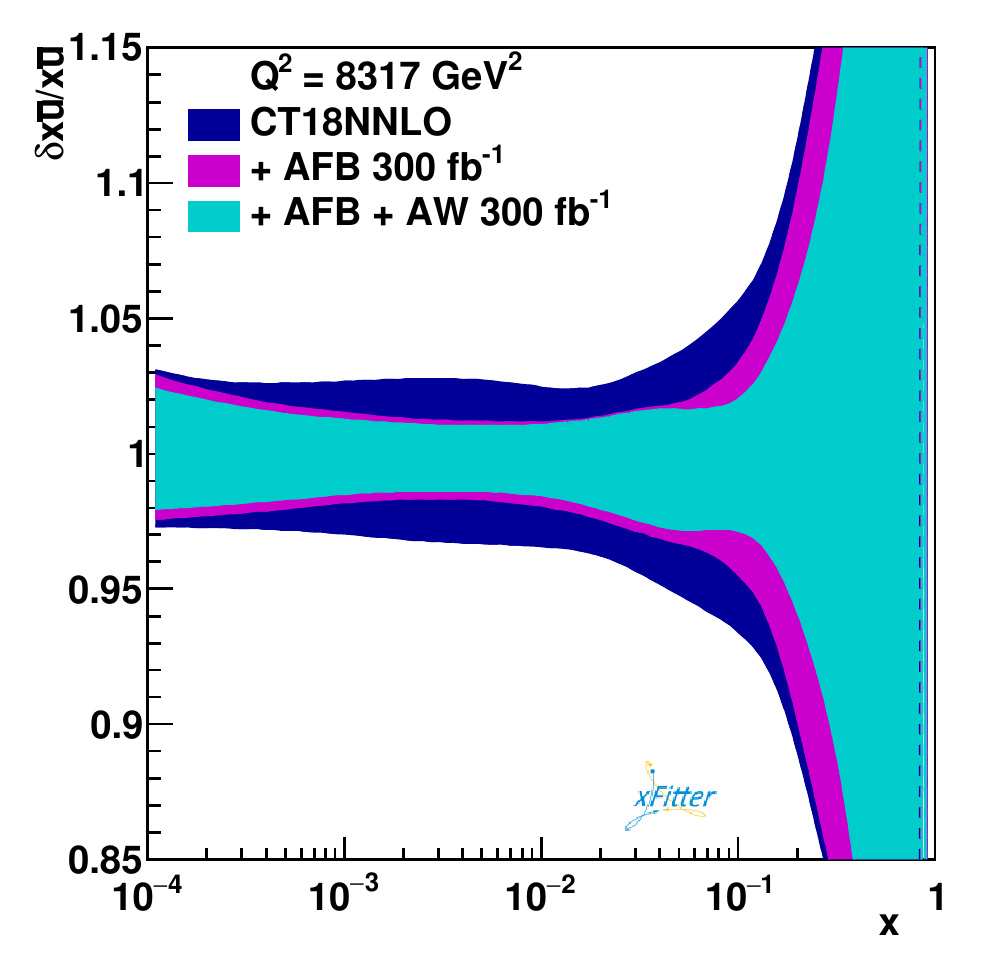}
\includegraphics[width=0.32\textwidth]{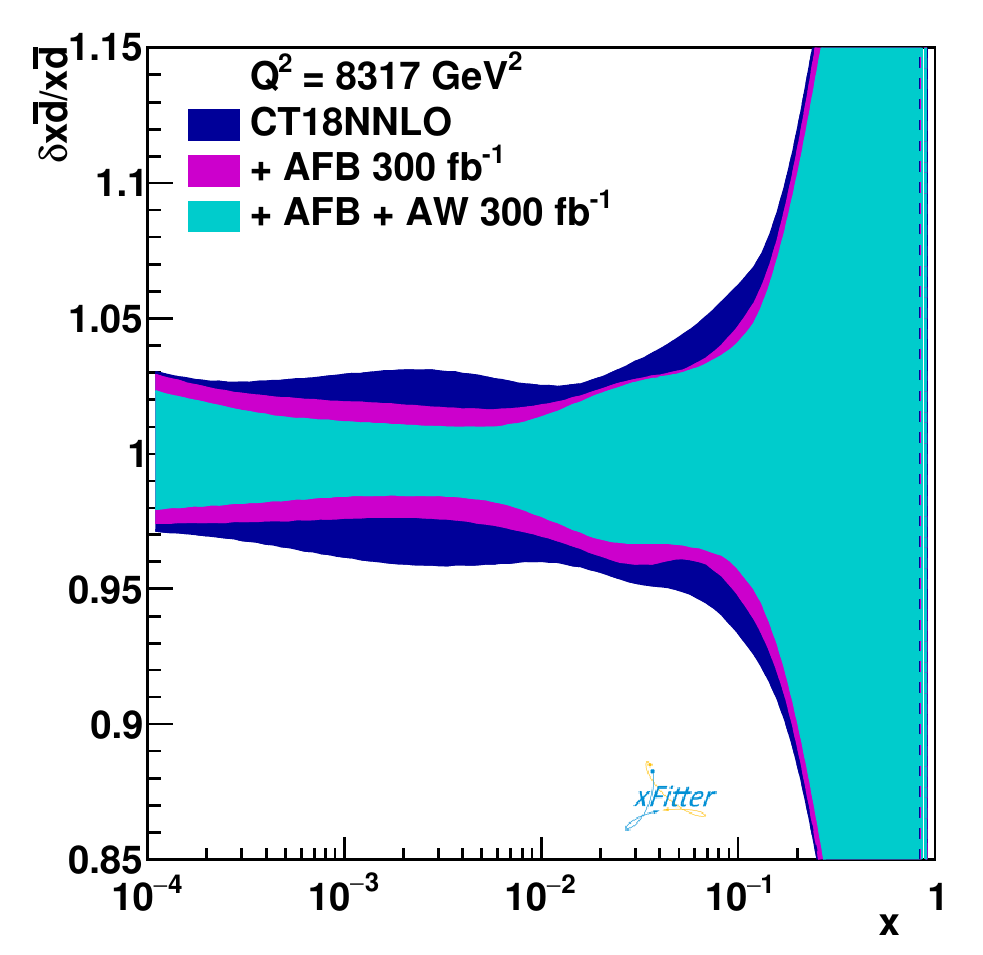}
\includegraphics[width=0.32\textwidth]{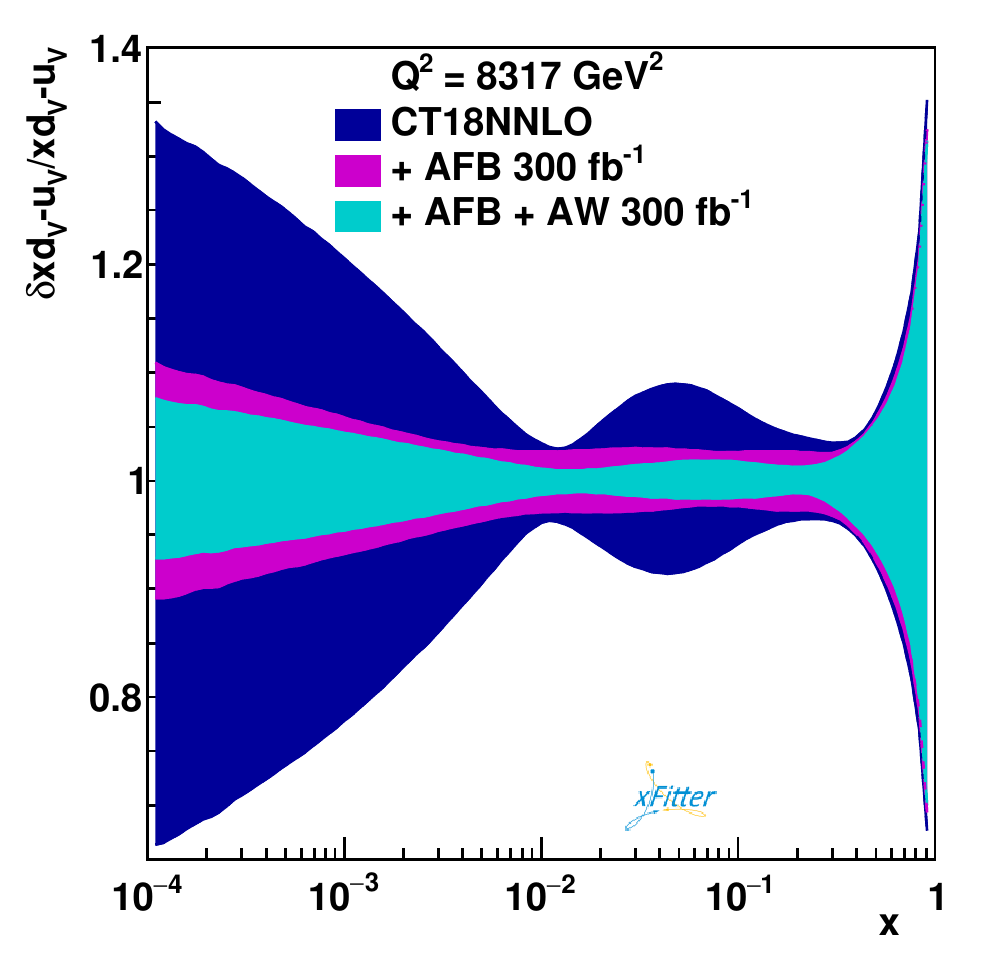}
\end{center}
\caption{Original CT18NNLO and profiled distributions of selected PDFs using either $A_{FB}$ or both $A_{FB}$ and $A_{W}$ pseudodata corresponding to an integrated luminosity of 300 fb$^{-1}$.}
\label{fig:PDF1}
\end{figure}
\section{Introduction}
Thanks to the large amount of data that is being and will be collected at Run-III and High Luminosity (HL) stage, precision measurements at the Large Hadron Collider (LHC) are reaching an unprecedented level of statistical accuracy, whilst Parton Distribution Functions (PDFs) uncertainties prevail. Drell-Yan (DY) lepton pair production provides some of the highest precision measurements at the LHC. The impact on PDFs from various observables related to the neutral DY process, such as the Forward-Backward Asymmetry ($A_{FB}$)~\cite{Accomando:2017scx,Accomando:2018nig,Accomando:2019vqt} or the longitudinal polarization coefficient $A_{0}$~\cite{Amoroso:2020fjw}, has been assessed, emphasising the constraining power of the clean di-lepton final state.\\
We study the impact of future measurements of forward-backward and lepton-charge ($A_W$) asymmetries in Drell-Yan (DY) events on PDF determination. Furthermore, the potential of combining $A_{FB}$ and $A_W$ to improve the determination of Standard Model (SM) observables, such as the mass of the $W$ boson $m_{W}$, as well as the sensitivity of specific beyond the SM (BSM) constructions featuring additional charged and neutral broad resonances in their spectrum is explored. This proceeding summarises the main results of Ref.~\cite{Fiaschi:2021okg}.

\section{$\mathbf{A_{FB}}$ and $\mathbf{A_{W}}$ complementarity}
\begin{figure}[t!]
\begin{center}
\includegraphics[width=0.32\textwidth]{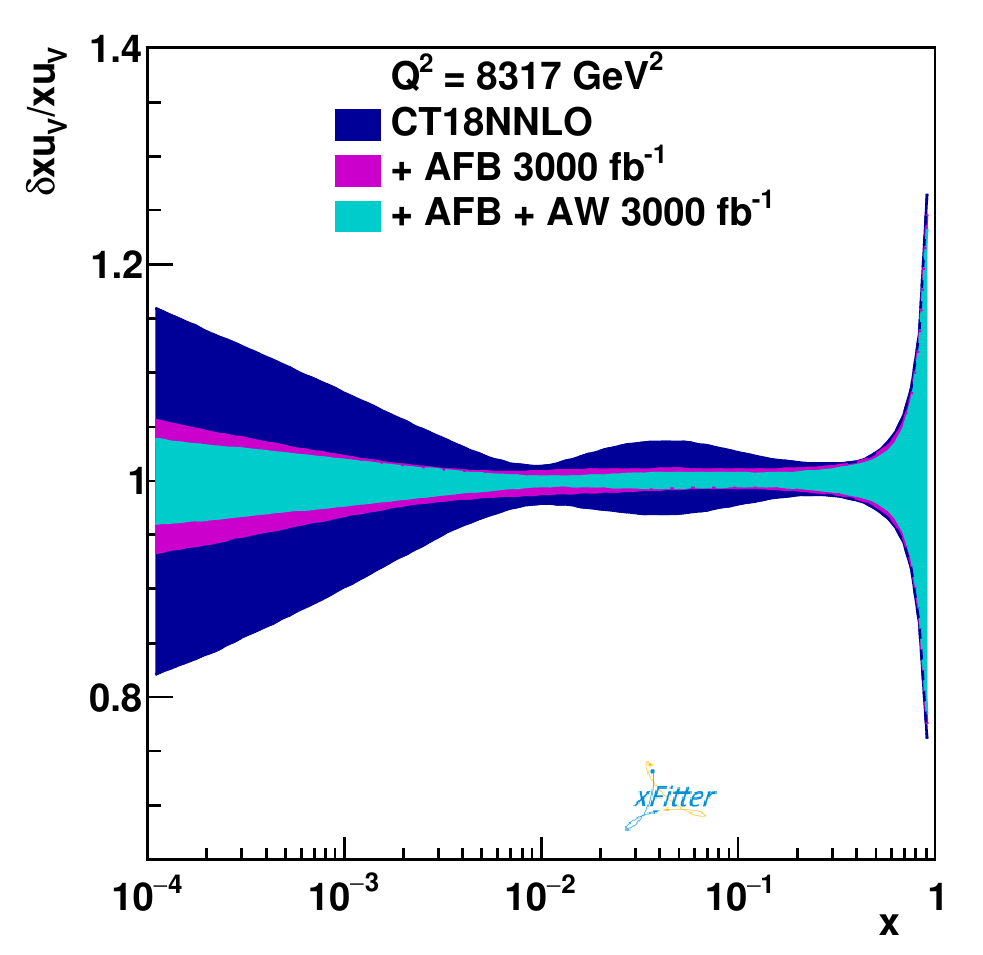}
\includegraphics[width=0.32\textwidth]{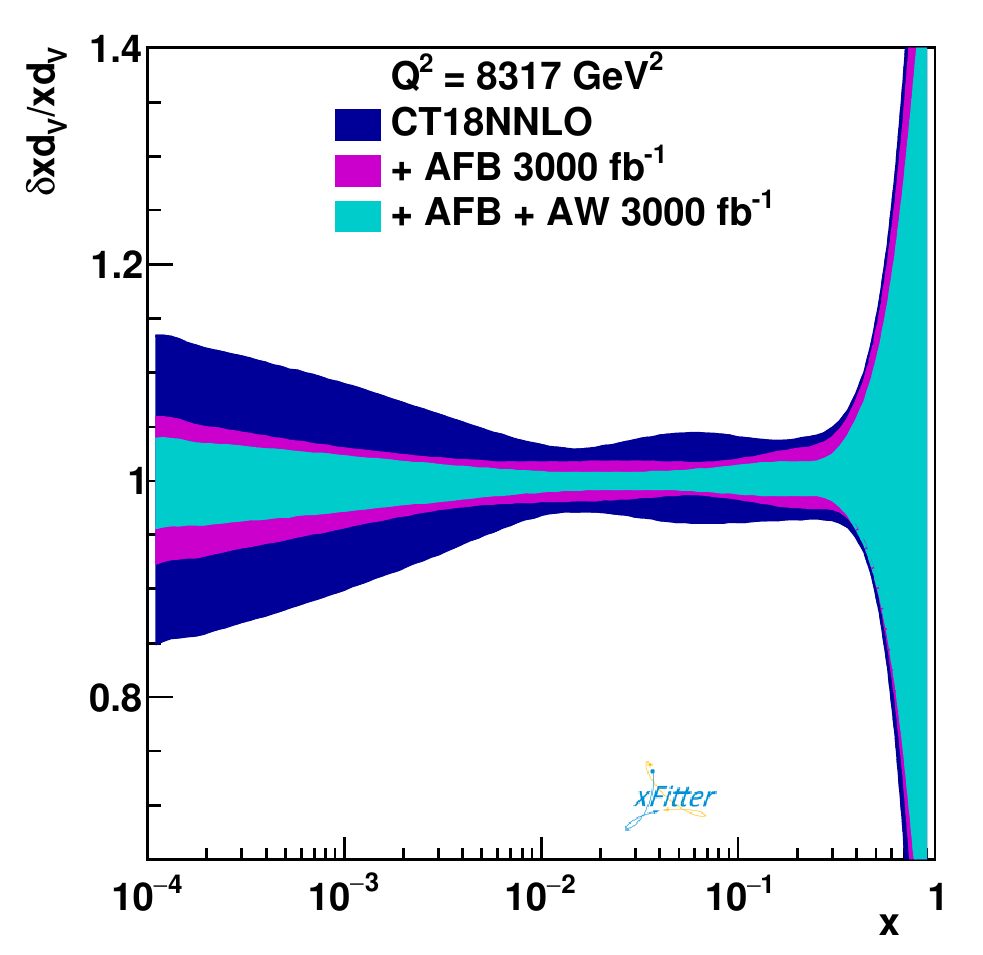}
\includegraphics[width=0.32\textwidth]{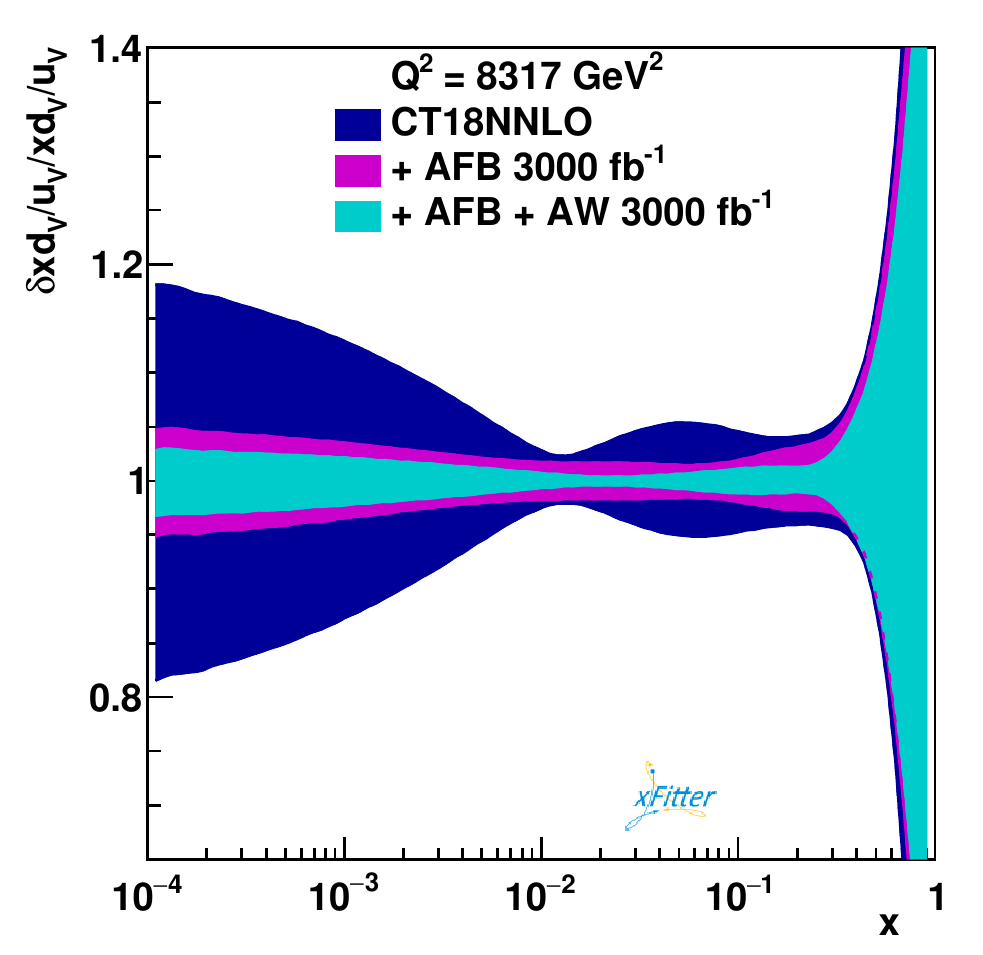}
\includegraphics[width=0.32\textwidth]{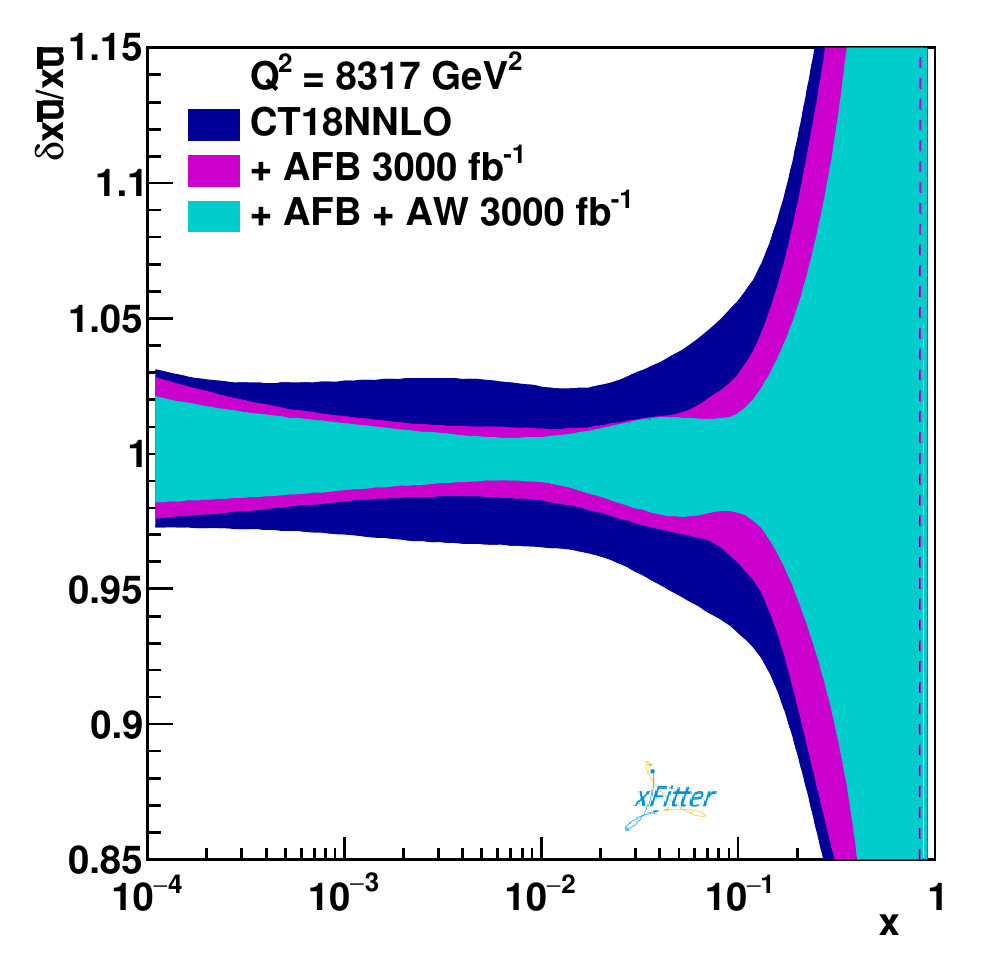}
\includegraphics[width=0.32\textwidth]{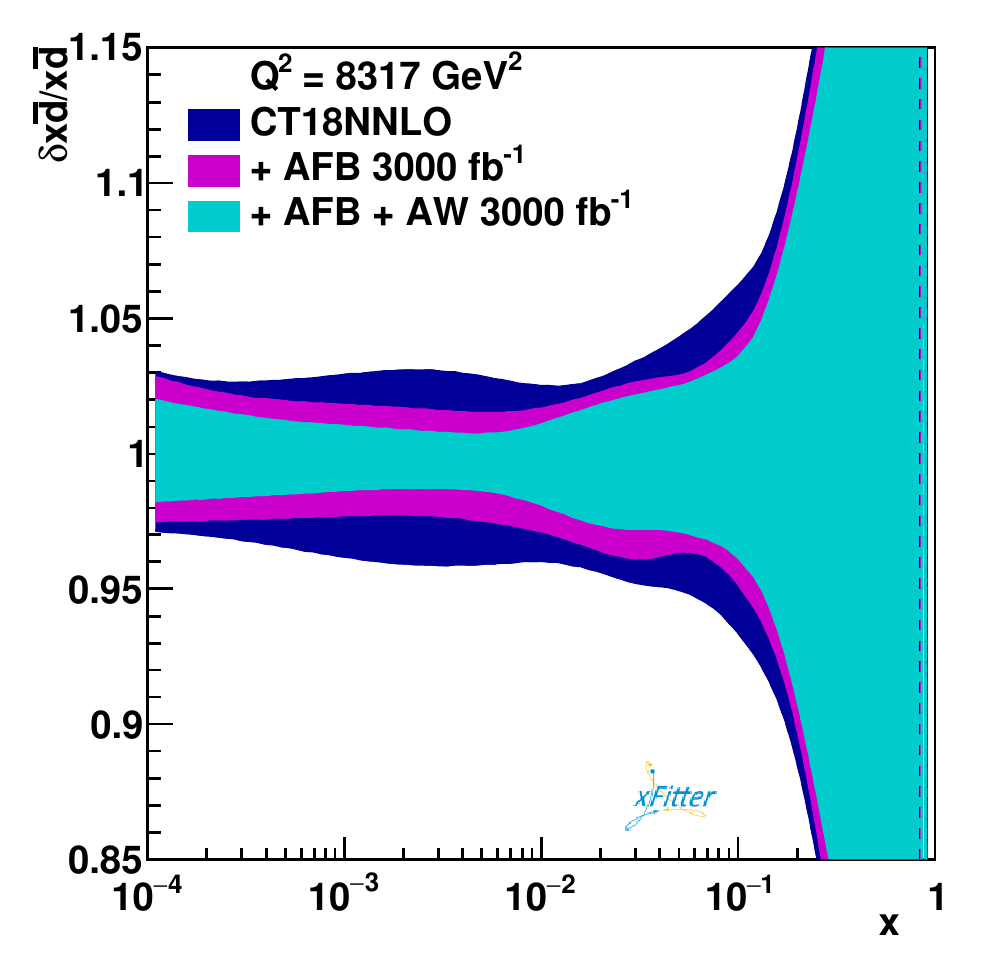}
\includegraphics[width=0.32\textwidth]
{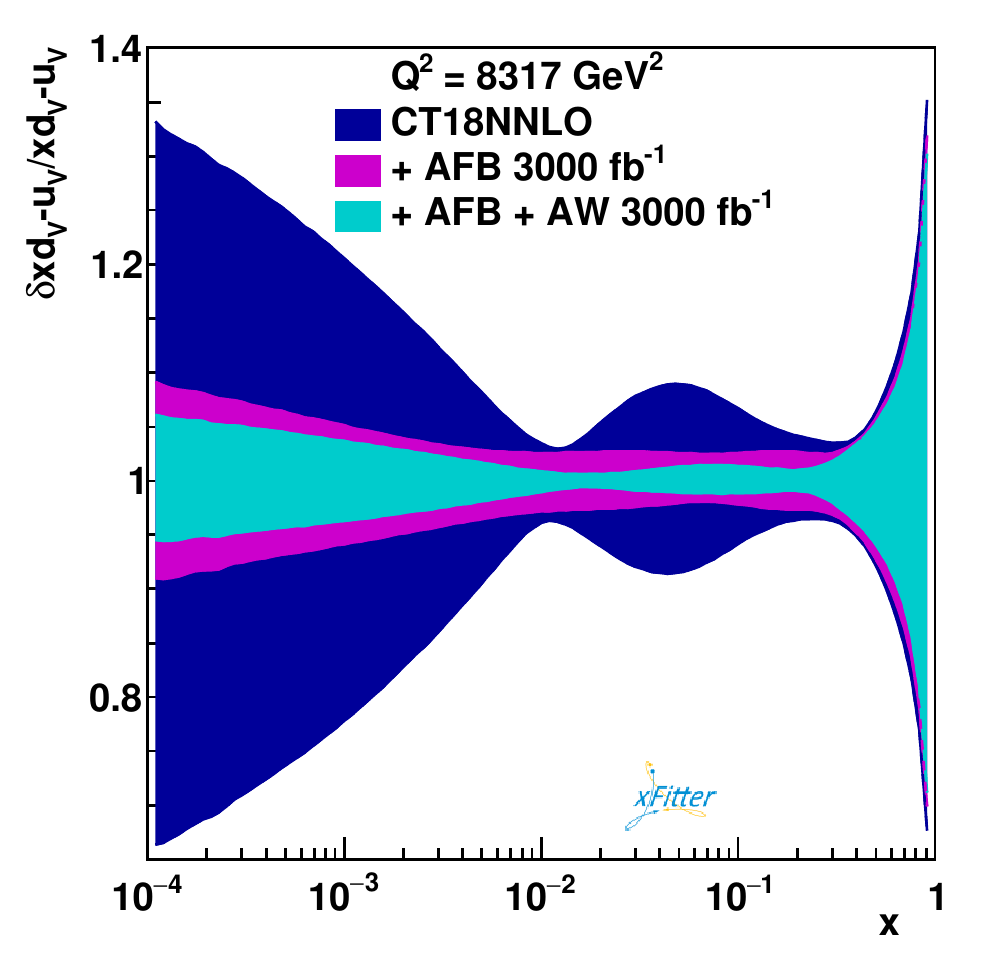}
\end{center}
\caption{Original CT18NNLO and profiled distributions of selected PDFs using either $A_{FB}$ or both $A_{FB}$ and $A_{W}$ pseudodata corresponding to an integrated luminosity of 3000 fb$^{-1}$.}
\label{fig:PDF2}
\end{figure}
In order to quantify the complementarity between $A_{FB}$ and $A_{W}$, the reduction of PDF uncertainties by profiling~\cite{Paukkunen:2014zia} the CT18 PDF set~\cite{Hou:2019efy} has been quantified using the \texttt{xFitter} framework~\cite{Alekhin:2014irh}. The $A_{FB}$ distribution as a function of the invariant mass of the dilepton pair in the final state ($m_{\ell\ell}$) has been implemented following the work done in Ref.~\cite{Accomando:2019vqt}. The implementation of $A_{W}$ as a function of the pseudorapidity of the charged lepton ($\eta_{\ell}$) has been similarly conducted~\cite{Fiaschi:2021okg}, with the inclusion of the fiducial cuts reported in Ref.~\cite{ATLAS:2019fgb}, using fixed-order (FO) next-to leading order (NLO) predictions computed with \texttt{MadGraph5{\_}aMC@NLO}~\cite{Alwall:2014hca} as well as next-to-next-to leading order (NNLO) QCD $k$-factors obtained from \texttt{DYNNLO}~\cite{Catani:2009sm}. We generated pseudo-data for two different integrated luminosity scenarios, namely 300 and 3000~fb$^{-1}$.\\
Figures~\ref{fig:PDF1} and~\ref{fig:PDF2} show the resulting profiled PDFs. The reduction in the PDF uncertainties bands is particularly visible in the valence distributions $u_{V}$ and $d_{V}$, as well as in their linear combination $x(d_{V} - u_{V})$. For values of Bjorken $x$ between 10$^{-3}$ and 10$^{-2}$, a moderate reduction of the PDF uncertainties can be observed for the sea distributions $\bar{u}$ and $\bar{d}$.

\section{Impact on SM measurements}
\begin{figure}[t!]
\begin{center}
\includegraphics[width=0.385\textwidth]{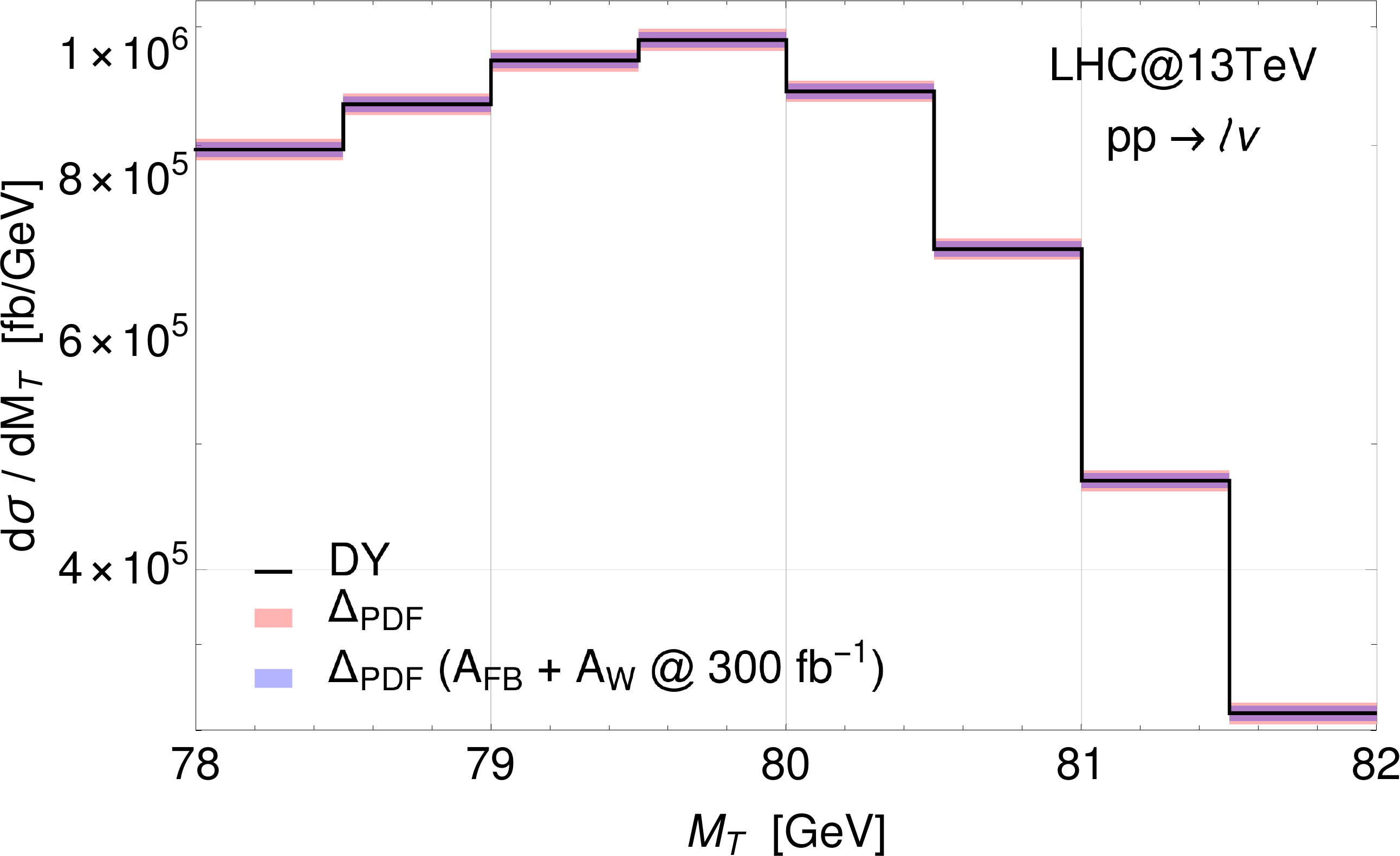}
\includegraphics[width=0.585\textwidth]
{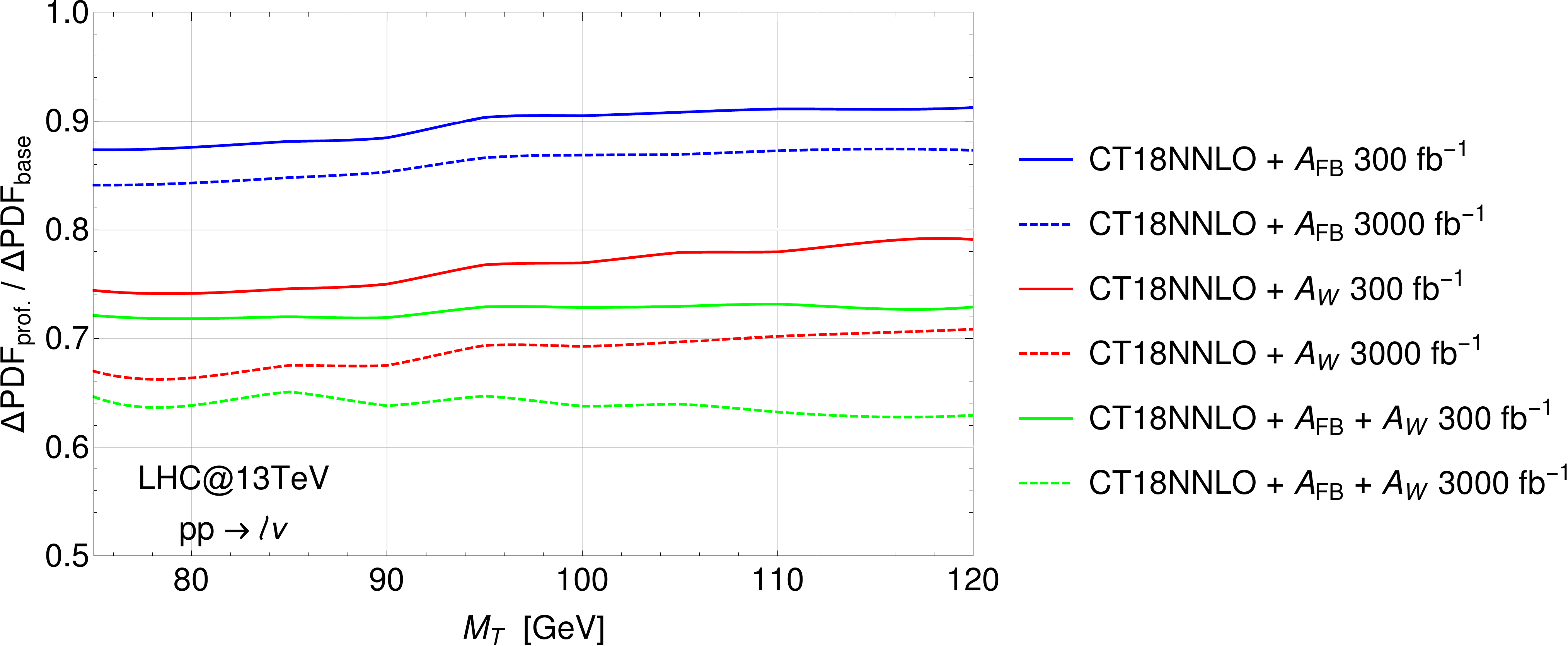}
\end{center}
\caption{Left: charged-current DY transverse mass distribution with PDF error. Right: relative improvement of PDF error on the transverse mass spectrum due to profiling based on $A_{FB}$, $A_{W}$ and their combination.}
\label{fig:mW}
\end{figure}
The above described PDF improvements could potentially impact precision measurements of SM parameters. For instance, determinations of the $W$ boson mass are carried out at the LHC from the lepton $p_{\mathrm{T}}$ distribution or from the $W$ transverse mass distribution, $m_{\mathrm{T}}^{W}$. This latter distribution is shown in Figure~\ref{fig:mW}, where its statistical and PDF uncertainty after the profiling can be found as well.\\
The curves in the right hand-side plot represent the relative improvement of PDF uncertainties after profiling using pseudo-data corresponding to different luminosities of $A_{FB}$ and $A_{W}$ and their combination. The PDF uncertainty would be reduced by about 26\% (43\%) using $A_{W}$ pseudodata at 300 (3000) fb$^{-1}$, by about 12\% (16\%) using $A_{FB}$ pseudodata at 300 (3000) fb$^{-1}$, and by about 28\% (46\%) combining the two pseudodata sets at 300 (3000) fb$^{-1}$.\\
Assessing the improvements on $m_{W}$ requires a refined analysis of normalised distributions (where reduction of PDF uncertainty is far more moderate) is beyond the scope of this work and we leave it for a forthcoming publication.

\section{Impact on BSM searches}
\begin{figure}[t!]
\begin{center}
\includegraphics[width=0.495\textwidth]{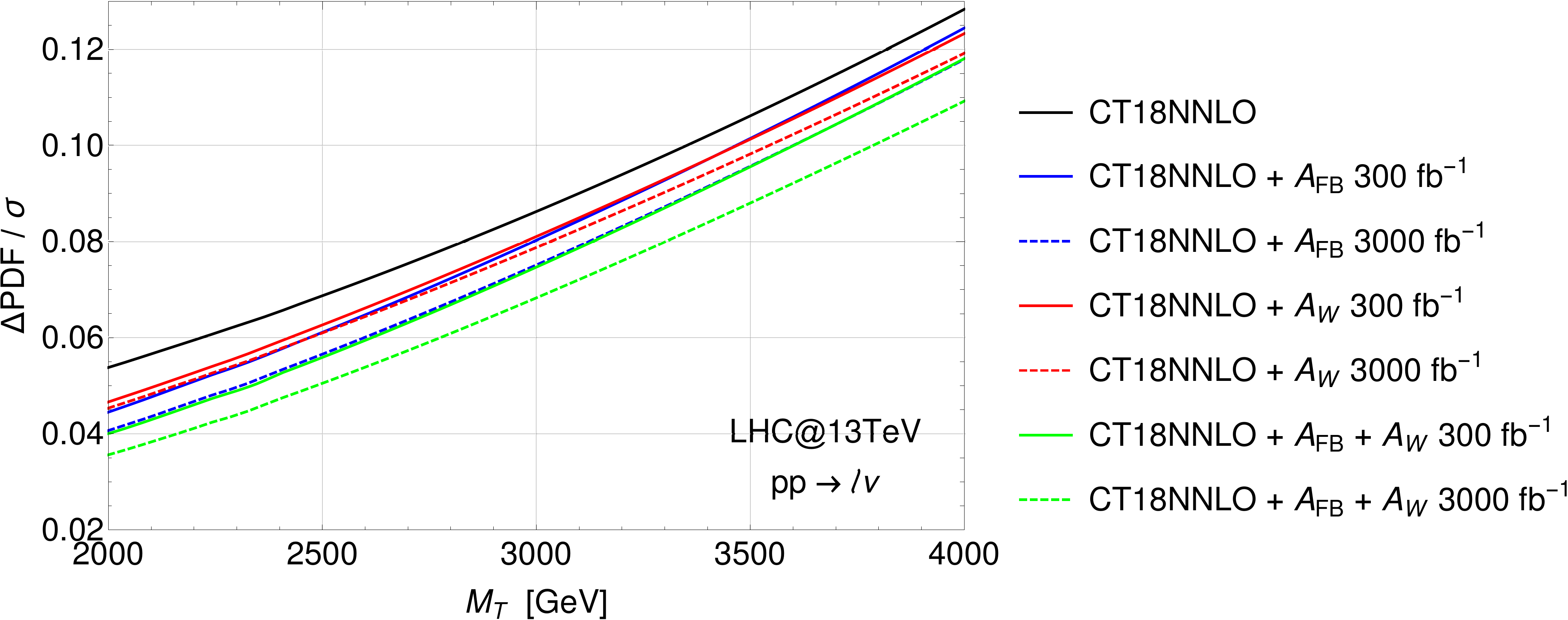}
\includegraphics[width=0.495\textwidth]{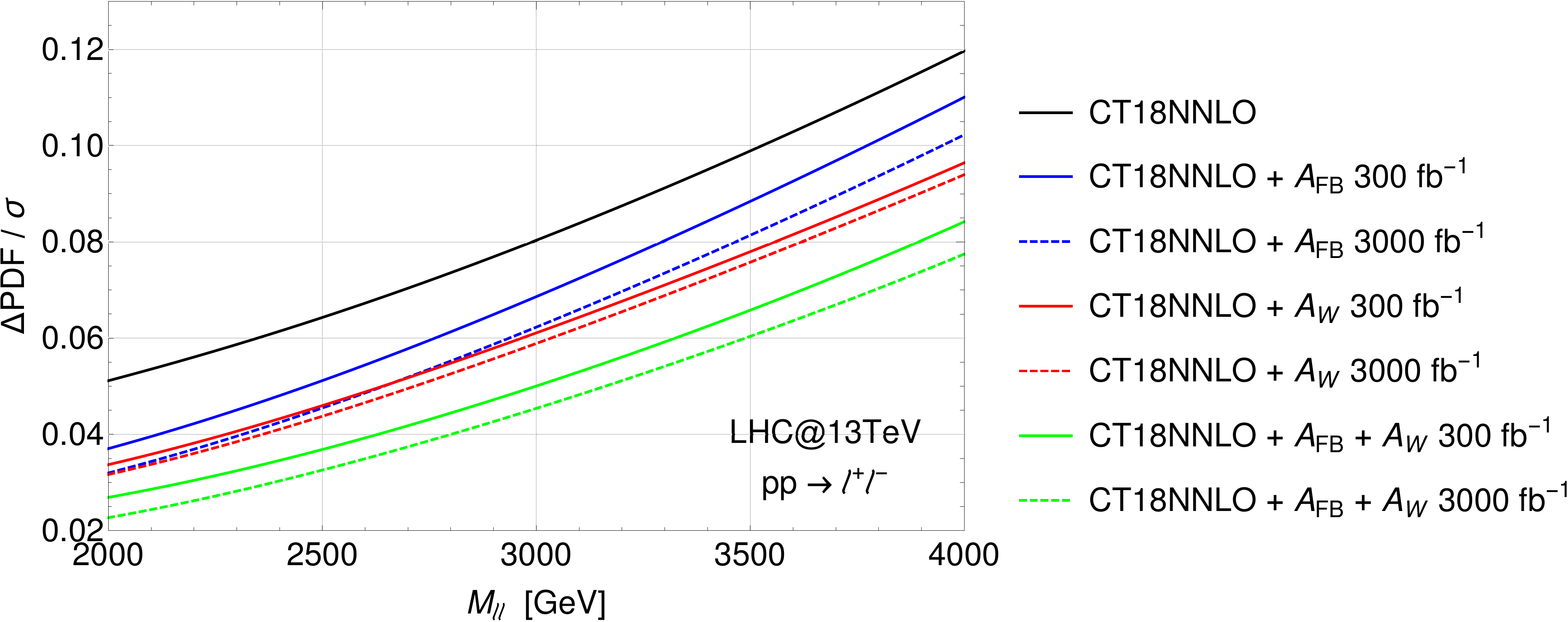}
\end{center}
\caption{Left: relative PDF error on the transverse mass spectrum of the charged DY channel. Right: PDF error on the invariant mass spectrum of the neutral DY channel.}
\label{fig:BSM}
\end{figure}
In the multi-TeV region e.g. at large invariant masses, the systematic uncertainties due to PDF determination is always a limiting factor for BSM physics searches. Generally, a Breit-Wigner (Jacobian) peak shape for the $Z^{'}$ ($W^{'}$) signal is assumed in standard analyses looking for heavy BSM resonances. If the resonance is broad, the usual bump hunt procedure is superseded by counting experiments, where an excess of events is sought over an estimated SM background expectation. In this context, the PDF uncertainty plays an important role in assessing the sensitivity to such BSM signals as its reduction would improve exclusion bounds or could lead to an early discovery~\cite{Accomando:2019ahs}.\\
The relative PDF uncertainty on the cross section of the original and profiled PDF sets using various sets of pseudodata is shown in Figure~\ref{fig:BSM}. The curves in the plot on the left hand-side are plotted as a function of the transverse mass in the charged DY channel, while in the plot on the right hand-side as a function of the di-lepton invariant mass in the DY neutral channel.\\
The combination of $A_{W}$ and $A_{FB}$ data further constrains the PDF relative uncertainty to 4.0\% (3.6\%) for a transverse mass of 2 TeV and to 11.8\% (10.9\%) for a transverse mass of 4 TeV when using an integrated luminosity of 300 (3000) fb$^{-1}$. The same combination further constrains the PDF relative uncertainty to 2.7\% (2.3\%) for an invariant mass of 2 TeV and to 8.4\% (7.8\%) for an invariant mass of 4 TeV when using an integrated luminosity of 300 (3000) fb$^{-1}$.

\section{Conclusion}
In this proceeding, the impact of the combination of the $A_{FB}$ and $A_{W}$ has been assessed by profiling the CT18 PDF set with pseudo data for various integrated luminosity scenarios, namely 300 and 3000 fb$^{-1}$. A significant reduction in PDF uncertainties for $u_{V}$ and $d_{V}$ valence quarks has been found for $x\gtrsim$ 10$^{-4}$, as well as a more moderate contraction in the antiquark $\bar{u}$ and $\bar{d}$ PDFs.\\
We have then found that the combined effect of $A_{FB}$ and $A_{W}$ leads to a 30\% improvement in the PDF uncertainties on the transverse mass in lepton-neutrino final states over a broad kinematic range measured at the LHC around the $W$ boson peak. Finally, we have investigated the impact of pseudo data from the peak region on the description of the multi-TeV region in both the neutral and charged DY channel. We have found that the constraints coming from the $A_{FB}$ and $A_{W}$ combination improve the relative PDF uncertainty by around 20\% in the invariant and transverse mass spectra, respectively, between 2 TeV and 4 TeV.

\end{document}